\documentclass[lettersize,journal]{IEEEtran}
\usepackage{amsmath,amsfonts}
\usepackage{algorithmic}
\usepackage{algorithm}
\usepackage{array}
\usepackage[caption=false,font=normalsize,labelfont=sf,textfont=sf]{subfig}
\usepackage{textcomp}
\usepackage{stfloats}
\usepackage{url}
\usepackage{verbatim}
\usepackage{graphicx}
\usepackage{cite}
\usepackage{comment}
\usepackage{xcolor}
\usepackage[colorinlistoftodos,prependcaption]{todonotes}
\usepackage{soul}
\usepackage{booktabs} 
\usepackage{siunitx}
\usepackage{chemist}
\usepackage[version=4]{mhchem}

\usepackage[british]{babel}

\begin{document}

\title{On Drug Delivery System Parameter Optimisation via Semantic Information Theory}

\author{\IEEEauthorblockN{Milica~Lekić, Mohammad~Zoofaghari, Ilangko~Balasingham,~\IEEEmembership{Senior Member,~IEEE}, Mladen~Veletić,~\IEEEmembership{Member,~IEEE}} 
\thanks{This work was supported in part by the Research Council of Norway (RCN: CIRCLE Communication Theoretical Foundation of Wireless Nanonetworks) under Grant 287112.}

\thanks{Milica Lekić is with the Department of Electronic Systems, Norwegian University of Science and Technology, 7491 Trondheim, Norway. e-mail: {milica.lekic@ntnu.no}}

\thanks{Mohammad Zoofaghari is with the Department of Electrical Engineering, Yazd University, Yazd 89195-741, Iran, and the Intervention Centre, Division of Technology and Innovation, Oslo University Hospital, 0372 Oslo, Norway. e-mail: zoofaghari@yazd.ac.ir}

\thanks{Ilangko Balasingham is with the Department of Electronic Systems, Norwegian University of Science and Technology, 7491 Trondheim, Norway, and the Intervention Centre, Division of Technology and Innovation, Oslo University Hospital, 0372 Oslo, Norway. e-mail: {ilangko.balasingham@ntnu.no}}

\thanks{Mladen Veletić is with the Intervention Centre, Division of Technology and Innovation, Oslo University Hospital, 0372 Oslo, Norway. e-mail: {mlavel@ous-hf.no}}

}
\markboth{IEEE TRANSACTIONS ON MOLECULAR, BIOLOGICAL, AND MULTI-SCALE COMMUNICATIONS,~Vol.~XX, No.~XX, XX~2025}%
{Shell \MakeLowercase{\textit{et al.}}: A Sample Article Using IEEEtran.cls for IEEE Journals}


\maketitle

\begin{abstract}
We investigate the application of semantic information theory to drug delivery systems (DDS) within the molecular communication (MC) framework. 
To operationalise this, we observe a DDS as a molecular concentration-based channel. Semantic information is defined as the amount of information required for a DDS to achieve its therapeutic goal in a dynamic environment. 
We derive it by introducing interventions, defined as modifications to DDS parameters, a viability function, and system-environment correlations quantified via the channel capacity.
Here, the viability function represents DDS performance based on a drug dose-response relationship.
Our model considers a system capable of inducing functional changes in a receiver cancer cell, where exceeding critical DDS parameter values can significantly reduce performance or cost-effectiveness.
By analysing the MC-based DDS model through a semantic information perspective, we examine how correlations between the internalised particle concentration $(Y)$ and the particle concentration in the extracellular environment $(X)$ evolve under interventions.
The final catalogue of results provides a quantitative basis for DDS design and optimisation, offering a method to determine optimal DDS parameter values under constraints such as chemical budget, desired effect and accuracy.
Thus, the proposed framework can serve as a novel tool for guiding DDS design and optimisation.
\end{abstract}

\begin{IEEEkeywords}
Semantic Information, Drug Delivery Systems, Molecular Communications.
\end{IEEEkeywords}

\section{Introduction}
 
\IEEEPARstart{D}{rug} delivery system (DDS) refers to the release and transport of excipients, which are tailored to physiological conditions and disease progression, to a specific site in the body in a controlled manner \cite{C4CC01429D}.
Despite substantial progress, the human body’s complexity and the numerous interacting phenomena within DDS present persistent developmental challenges. A key unresolved issue in moving away from a one-size-fits-all approach is the development of optimised, patient-specific DDS \cite{EZIKE2023}. The optimisation is characterised by (at least) one of the following criteria: ‘precision targeting’, ‘controlled release’, ‘maximised therapeutic efficacy’, ‘minimised side-effects’, ‘enhanced patient compliance’, ‘cost-effectiveness’, and ‘adaptability and personalisation’. This challenge stems largely from our current inability to sufficiently quantify DDS effects, which impedes precise design and optimisation and limits the potential for personalised treatments.

The integration of computational methods in drug delivery has proven to minimise the need for time-consuming and costly in vitro and in vivo experiments \cite{Begum2021, Wenbo2022, Hm2013, Youyong2010}. However, limitations imposed by the complexity of accurately measuring and interpreting drug concentrations and effects over time, as well as the variability in biological systems, highlight the need for a shift towards the analysis of the therapeutic effect quantification.

From the perspective of information theory, drug delivery can be interpreted as an information delivery process that is governed by the fundamental principles of transmission, propagation, and reception focusing on efficiently and accurately transferring data as a 'message' through a medium to a specific target. Applying this framework to drug delivery, the drug administration system can be viewed as the sender, while the target cells can be viewed as the receiver.
Recognising this similarity, we can formulate drug delivery problems under the paradigm of molecular communication (MC) \cite{CHAHIBI2017}, where molecules serve as information carriers, modulating cellular responses and physiological processes. This concept forms molecular information which refers to the biochemical or biophysical data encoded, transmitted, and processed at the molecular level within biological or synthetic systems.

On one hand, MC has been partly successful in modelling complex DDS by considering body conditions and drug parameters through electrical circuit representations \cite{CHAHIBI2013}, stochastic models with queuing systems \cite{WYSOCKI2013}, information theoretical analysis \cite{Michelusi2015}, and the abstraction of DDS through a layered architecture \cite{DONG2016}. 
On the other hand, there has been limited progress in addressing the problem of finding the ideal design of the DDS parameters under constraints such as the chemical budget, desired effect and accuracy \cite{CHAHIBI2017}. This has prevented the field of MC from providing significant complementary knowledge towards optimising drug delivery and personalised medicine.

A limitation in current MC approaches lies in the inappropriate abstraction level associated with information encoded into molecular messengers \cite{brand_semantic_2024}. Information transmission occurs at multiple levels of abstraction: (i) accurate syntactic information (technical level), (ii) message meaning/significance (semantic level), and (iii) actions induced by the message (effectiveness level) \cite{brand_semantic_2024}. Syntactic information quantifies various statistical correlations between two systems without considering their meaning. Shannon’s well-known information theory studies syntactic information by providing measures that quantify how much knowledge of one system’s state reduces statistical uncertainty about another system’s state. Shannon focused on the engineering problem of accurately transmitting messages across a telecommunication channel, explicitly avoiding questions about the messages’ meaning. 

Conversely, semantic information refers to information that is meaningful for a system rather than merely correlational. For example, when a frog sees a small black spot in its visual field, it instinctively extends its tongue to capture what it perceives as a fly \cite{kolchinsky_semantic_2018}. This stimulus–response behaviour was induced because the small black spots in the visual field indicated the presence of flies, which are a valuable food source for the frog’s survival. Therefore, the small black spot in the visual field of a frog has semantic information and signifies the likely presence of flies.
As illustrated here, living systems possess intrinsic goals such as maintaining the living state, viability, by adapting their behaviour in response to different environments \cite{bartlett2025}. This suggests that living organisms actively use information for functional purposes, which supports the application of the semantic information perspective to DDS as a living system.
Furthermore, as stated in \cite{brand_semantic_2024}, MC systems are characterised by relatively poor performance on the syntactic level (low achievable data rates, high bit error rates, etc.) \cite{GURSOY2022} and good performance on the semantic and effectiveness level (complex adaptations to environmental conditions) \cite{Pál2017}, indicating that semantic information may be the most relevant measure to consider in MC systems \cite{Malcolm2023, ChengYukun2024}, including particle-based drug delivery.

Carnap and Bar-Hillel first introduced semantic information theory to quantify the semantic information encoded in a sentence of a particular language \cite{Carnap1952}. However, its application is limited to language studies and not applicable to physical scenarios. In contrast, Kolchinsky and Wolpert proposed a theory of semantic information grounded in the intrinsic dynamics of a physical system and its environment \cite{kolchinsky_semantic_2018}. It builds upon ideas from Shannon’s information theory and non-equilibrium statistical physics.

We hypothesise that by applying the concept of semantic information to drug delivery, it becomes possible to identify portions of DDS parameters that are effectively ‘meaningful’ and ‘meaningless’ in terms of their impact on a target cell. 
For example, in an overdosed scenario a ‘meaningless’ portion of the drug refers to that part of the administered substance which, if reduced or eliminated at the target cell, would not alter the therapeutic effect significantly. This means that despite its presence, this portion does not contribute to the desired changes in the target cells (e.g., altering behaviour or function, or inducing death in harmful cells). This would, for example, reveal opportunities to reduce the chemical dosage while maintaining therapeutic efficacy, further leading to improvements in chemical budget optimisation and ‘cost-effectiveness’.

The theory of semantic information by Kolchinsky and Wolpert has been recently demonstrated for the information-theoretic analysis of bacterial chemotaxis --- a natural MC system \cite{brand_semantic_2024}. Chemotactic bacteria detect and follow nutrient gradients in their environment with the goal of adapting and surviving. The observed semantic information reveals how much information acquisition of the bacteria can be intervened without seriously reducing its chances of survival.
The same concept has been applied to the design of synthetic cells as an MC-based system \cite{ruzzante_synthetic_2023, maurizio2}. Synthetic cells process environmental signals emitted by cancerous cells in order to produce therapeutic agents targeting the cancer effectively. The computed semantic information reveals the amount of information that is causally necessary for a synthetic cell to complete its task.
These studies utilise the semantic information concept to describe dynamic information exchange between a system and its environment across different contexts. However, we present its first application to optimisation and design, specifically focusing on optimising and designing the drug delivery process through DDS parameters. Unlike these works, we extend semantic information measures beyond the originally proposed mutual information and transfer entropy \cite{kolchinsky_semantic_2018} and define a viability function which is rooted in a biochemical MC system suitable enough for a DDS analysis.

In this paper, we consider a molecular concentration-based channel model as a DDS consisting of a particle release point and a cancer cell modelled as a reactive receiver \cite{pic}. This model captures chemotherapeutic particle delivery dynamics in anticancer treatment, where the channel impulse response represents the detection probability of the released particle. Building on this foundation, we develop the semantic information framework, rooted in \cite{kolchinsky_semantic_2018}, for the considered computational MC-based DDS model to discover the ‘meaningful’ and ‘meaningless’ portions of a subset of DDS parameters that can be altered.
This approach quantifies the drug delivery performance on the semantic level based on the particle internalisation probability by a cancer cell and the dynamic information exchange defined by the system parameters.
Finally, the results form a catalogue that provides a quantitative basis for the DDS design and optimisation, offering a method to extract optimal DDS parameter values under constraints such as chemical budget, desired effect, and accuracy.

The remainder of this paper is organised as follows. Sections II and III introduce the considered MC DDS model and the proposed semantic information framework, respectively. Numerical results and discussion are presented in Section IV, and Section V concludes the paper and outlines topics for future work.


\section{System Model}\label{sec:2}
The DDS is modelled as a particle intensity channel (PIC), that is described by the main characteristics of particle transmission and uptake processes within an MC framework. In the PIC model, the input $X$ is the probability of particle release, which is continuous over the interval $\left\lbrace0,1\right\rbrace$. The channel output $Y$ is the number of released particles that are counted by the receiver (Rx) during the symbol interval $\tau$. The particles travel independently from the transmitter (Tx) to the Rx via diffusion, characterised by the diffusion coefficient $D$, from areas of high concentration to areas of low concentration due to random molecular motion. The output $Y$ can be modelled using the binomial distribution, where the probability of observing $y$ successfully received particles for the maximum number of released particles, $N$, is given by:
\begin{equation}\label{eq: binomial_distributon}
P(Y = y \vert X = p) = \binom{N}{y} \left(pP_\mathrm{i}\right)^y\left(1-pP_\mathrm{i}\right)^{N-y}.
\end{equation}
A single particle is selected to be released by the Tx with probability $p$. The probability $P_\mathrm{i}(\tau \vert \overrightarrow{r_0})$ \cite[eq. (20)]{pic} denotes the probability that a given particle released at $\overrightarrow{r_0} = (x_0, y_0, z_0)$ and time $t_0 = 0$ is internalised into the Rx, where the actual internalisation action, i.e., detection can happen any time during the symbol interval $\tau$. 
We observe the case when the entire surface of the receiver is fully covered with infinitely many receptors and that multiple particles can react with the same receptor \cite[eq. (16)]{pic}.
We refer to $P_\mathrm{i}(\tau \vert \overrightarrow{r_0})$ as the detection probability or channel impulse response of the system. The channel is used in a time-slotted manner, meaning that at the start of each time slot, the Tx can release $N(\lambda, \tau) = \lfloor \lambda \tau \rfloor$ particles, where $\lambda$ represents the particle release rate with units $[\lambda]= \si{particle\,\second^{-1}}$.

The channel impulse response of the system $P_\mathrm{i}(\tau \vert \overrightarrow{r_0})$ operates in a homogeneous 3D fluid environment with constant physical and chemical properties \cite{pic}. The Tx is modeled as a point source that releases particles simultaneously in bursts, while the Rx is a spherical structure with a fixed radius $a$, located at the centre of the coordinate system. The particles are identical and indistinguishable upon release. 
Particles in the medium can undergo: 1) degradation, where they decay over time at a rate $k_\mathrm{d}$, 2) reversible binding to receptors on the Rx surface, characterised by the binding $k_\mathrm{f}$ and recycling $k_\mathrm{b}$ rate constants, and 3) internalisation into the Rx at a rate $k_\mathrm{i}$ leading to detection.

The considered PIC model effectively describes particle transmission and uptake in a DDS, where therapeutic agents are released from a source and absorbed by target cancer cells, aligning with real-world DDS. The model accounts for the random motion of drug particles via diffusion, which mirrors real drug transport in the bloodstream or interstitial fluid. The interaction between drug molecules and target cell receptors follows biochemical reaction kinetics, capturing key pharmacokinetic and pharmacodynamic processes. The model includes particle degradation, a critical factor in DDS, ensuring realistic consideration of drug stability and efficacy over time. The channel impulse response captures the probability of successful drug uptake, making it a relevant performance metric for optimising DDS design.

\section{Proposed Semantic Information Framework}\label{sec:3}
In this section, we first outline the general semantic information framework for physical non-equilibrium systems, rooted in \cite{kolchinsky_semantic_2018}. Building on this foundation, we develop a semantic information framework tailored to the MC DDS model.

\subsection{Semantic Information Theory}\label{sec:3a}
Semantic information is formally defined as the subset of syntactic information that a system has about its environment that is causally necessary for the system to achieve its goal or maintain its existence over time \cite{kolchinsky_semantic_2018}. This definition relies on the specification of a viability function, the formulation of so-called counter-factual interventions (intervened distributions), and the use of a syntactic information measure. A flexible manner has been suggested in defining these factors allowing them to be chosen in a relevant way to the dynamics of the system under study.

A \textit{viability function} $V(\tau)$ quantifies the system’s ‘degree of existence’ at a given time $\tau$ as its ability to remain in a low entropy state \cite{kolchinsky_semantic_2018}. In drug delivery, an optimal system ensures controlled release of therapeutic molecules, as well as their targeting toward recipients. There is often a naturally defined ‘viability set’, which is a set of states in which the system $\mathcal{Y}$ can continue to perform self-maintenance functions. Typically, the viability set is a very small subset of the overall state space $Y$. Given a particular viability set $\mathcal{A} \subseteq Y$, a natural definition of the viability function is the probability that the system’s state is in the viability set, $P(Y_\mathrm{\tau} \in \mathcal{A})$ \cite{kolchinsky_semantic_2018}. This definition necessitates the specification of the viability set.

A \textit{counter-factual intervention} refers to any action applied to one part of a system or environment that results in changes in other parts of the system or environment \cite{kolchinsky_semantic_2018}. Here, counter-factual interventions may be associated with any parametric change that characterises drug delivery, for example, the target specificity of active substances. A counter-factual intervention may lead to one of two scenarios, depending on the difference $\Delta V$ between the system’s viability after time $\tau$ under the actual distribution $V_{\mathrm{a}}(\tau)$ versus the system’s viability after time $\tau$ under the intervened distribution $V_{\mathrm{i}}(\tau)$ $(\Delta V = V_{\mathrm{a}}(\tau) - V_{\mathrm{i}}(\tau))$.
The term actual distribution denotes the original distribution of the joint system–environment trajectories over time $\tau$, which defines the timescale of interest \cite{kolchinsky_semantic_2018}. The possible scenarios are:

1) A positive difference $\Delta V(\tau) > 0$ means that at least some of the syntactic information between the system and its environment plays a causal role in drug effectiveness. In that case, the system’s viability is not increased but the counter-factual intervention, apart from the optimal intervention, scrambles some of the syntactic information. Under the optimal intervention, none of the syntactic information between the system and its environment is scrambled. In other words, the viability of the system under actual distribution and counter-factual distribution is the same, but the amount of syntactic information is the smallest. Thus, the amount of semantic information possessed by the system is the amount of syntactic information preserved by the optimal counter-factual intervention. 
In terms of drug delivery, we can decompose the total amount of syntactic information of DDS efficiency into ‘meaningful part’ (the semantic information) and ‘meaningless part’ (the rest) and seek to optimise drug delivery excluding the ‘maximised therapeutic efficacy’ criterion.

2) A negative difference $\Delta V(\tau) < 0$ indicates that some of the syntactic information between the system and its environment is not relevant to drug effectiveness and may even hinder the therapeutic outcome. This also suggests that certain elements of the syntactic information are ‘noisy’ or counterproductive in the context of drug delivery. In this case, the intervened distribution leads to better viability for the system than the actual distribution, and we seek to optimise drug delivery including the ‘maximised therapeutic efficacy’ criterion.

Finally, semantic information is linked to a syntactic information measure but is not tied to any particular. 
Since the system under study is a (molecular) communication system, we propose the maximal mutual information, i.e., channel capacity as an elegant quantitative measure to derive semantic information received by the physical system influenced by an external environment in the initial distribution.

Notably, the computation of the semantic information is strictly dependent on the time instant $\tau$ at which the viability and channel capacity are computed \cite{kolchinsky_semantic_2018}. Furthermore, the structural and dynamical aspects of underlying system model also play a crucial role in shaping semantic information \cite{Kuramoto}.

\subsection{Semantic Information in Drug Delivery System}\label{sec:3b}

The system under consideration in this paper is a drug delivery composed of a target cancer cell system $\mathcal{Y}$ with a state space $Y$ described by physical variables $y$, and its environment $\mathcal{X}$ with a state space $X$ described by physical variables $x$.
Computing semantic information would then mean identifying a portion of the syntactic information within the target cell that is causally necessary to achieve a specific therapeutic goal. 
This goal can be achieved by delivering a sufficient number of chemotherapeutic particles or accommodating system parameters. 
Alterations in the environment $\mathcal{X}$ influence the dynamics of the system $\mathcal{Y}$, i.e., the dynamical system $\mathcal{Y}$ collects information about $X_{\tau}$ and adapts its behaviour accordingly \cite{brand_semantic_2024}.
According to the system model definition in Sec. \ref{sec:3a}, the environment $\mathcal{X}$ is characterised by the following system parameters: the diffusion coefficient $D$, the release rate of particles $\lambda$, the release point $r_\mathrm{0}$, the receiver radius $a$, the degradation rate $k_\mathrm{d}$, and the forward $k_\mathrm{f}$, backward $k_\mathrm{b}$, and internalisation $k_\mathrm{i}$ chemical reaction rate constants.
The system $\mathcal{Y}$ is associated with the internalised level of the considered molecules, where the cancer cell's internalisation capability is limited by the aforementioned environment characteristics, while the environment $\mathcal{X}$ is associated with the molecule concentration in the extracellular environment.

To compute the semantic information of the considered model, we first make use of a scalar-valued viability function $V(\tau)$ that quantifies the degree of existence of the system $\mathcal{Y}$. Further, we consider different counter-factual interventions that scramble the correlation between $\mathcal{Y}$ and $\mathcal{X}$ in the initial configuration. To quantify the interaction between the system $\mathcal{Y}$ and environment $\mathcal{X}$ after an intervention is applied, we specify the syntactic information measure. Finally, we conclude the system model analysis with a measure of semantic information at time $\tau$ between the system and environment.

\subsection{Viability Function}\label{sec:3c}
The viability function $V(\tau)$ can be described as a real-valued function that quantifies the degree of existence of the system $\mathcal{Y}$ over time and/or under specific conditions, with multiple possible definitions \cite{kolchinsky_semantic_2018}.
For example, the viability function $V[k]$ is defined as the percentage of simulations for which a chemotactic bacterium was still alive at time step $k$ \cite{brand_semantic_2024}. Alternatively, the viability function can be set as the negative of the Shannon entropy $S$ of the marginal distribution $p(Y_\mathrm{t})$ over the states of the system $\mathcal{Y}$ at time $t$ \cite{kolchinsky_semantic_2018, ruzzante_synthetic_2023}. Other formulations include the expected system lifetime in a foraging model \cite{Sowinski2023} and the expected proportion of the planet's habitable area covered by a biome \cite{sowinski2024}.

As there are numerous ways to formulate the viability function, we draw inspiration from real DDS applications and conceptualise $V(\tau)$ as a standard measure of DDS performance based on drug dose-response \cite{HACKER2009, zimmer2016, roman2016}. In the context of our study model, the viability function also quantifies the ‘health’ of the target cancer cell, which naturally creates an environment conducive to its survival. 
In practical DDS applications, viability encompasses factors such as caner cell viability, healthy cell toxicity, side effects, systemic toxicity, microenvironmental factors, and drug retention and clearance.
To ensure our model captures the essential aspects of the drug delivery process, we define viability in terms of DDS parameters that directly impact therapeutic outcomes. Accordingly, we model the viability function as a variation of the Hill equation \cite{hill1910}, commonly applied in pharmacology to describe drug dose-response relationships:
\begin{equation}\label{e:V1}
    V(\tau) = \dfrac{1}{1 + \left( \dfrac{C_{\mathrm{int}}(\tau)}{C_{\mathrm{th}}} \right)^{n}},
\end{equation}
where $C_{\mathrm{int}}(\tau)$ represents the internalised particle concentration per cell at time $\tau$, $C_{\mathrm{th}}$ is the threshold concentration for 50\% viability reduction, and $n$ is the Hill coefficient controlling the steepness of the viability curve. This form of the Hill equation ensures that viability decreases with increasing internalised particle concentration.
We define $C_{\mathrm{int}}(\tau)$ using the probability that the released particles are internalised by a caner cell:
\begin{equation}\label{e:c}
    C_{\mathrm{int}}(\tau) = P_{\mathrm{i}}(\tau \vert r_\mathrm{0}) N(\lambda, \tau).
\end{equation}
We use a notation for $C_{\mathrm{int}}(\tau)$ without explicitly indicating its dependence on $\lambda$ and space.
This formulation captures the release rate of particles $\lambda$ and the chemical constant rates for binding $k_\mathrm{f}$, unbinding $k_\mathrm{b}$, internalisation $k_\mathrm{i}$, and degradation $k_\mathrm{d}$. 
Therefore, the viability function reflects the inherent coupled dynamics between the system and environment \cite{Sowinski2023}.
It also reflects whether the system $\mathcal{Y}$ can maintain its self-sustaining functions, as $C_{\mathrm{int}}(\tau)$ provides an upper bound on the probability of particles that can be concentrated within a cancer cell. This definition of $V(\tau)$ enables its adaptation to specific drugs and DDS types, as it is based on the PIC approach.

\subsection{Counter-factual Interventions}\label{sec:3d}
To quantify semantic information, we conduct a series of controlled counter-factual interventions by altering specific drug delivery parameters to scramble the initial correlation between $\mathcal{Y}$ and $\mathcal{X}$ (at $t = 0$). The viability function and the semantic information measure then identify which correlations are essential for achieving the therapeutic goal.
These interventions induce the transformation of an initial distribution into an intervened distribution. 
We define the initial distribution by the default set of system parameters (Table \ref{tab:tab}).
By considering different interventions, we observe a trade-off between the amount of preserved syntactic information versus the resulting viability of the system at time $\tau$, represented by an information/viability curve \cite{kolchinsky_semantic_2018}.

As interventions, we observe the modification of 1) the drug dosage, expressed in the release rate $\lambda$, 2) the delivery performance, partially affected by the degradation reaction rate constant $k_\mathrm{d}$, and 3) the targeting specificity, defined by the binding, unbinding, and internalisation reaction rate constants, $k_\mathrm{f}$, $k_\mathrm{b}$, and $k_\mathrm{i}$, respectively.
We base the selection of DDS parameters as counter-factual interventions, among those considered in the system model (see Sec. \ref{sec:3a}), on the fact that they can be altered, i.e., that it is feasible to adjust them through the process of drug design and characterisation.
We select the parameter values to induce a positive $\Delta V$ (the difference of the viability function between the initial distribution and the counter-factual interventions).

\subsection{Semantic Information Measure}\label{sec:3e}
Transfer entropy is a prevailing syntactic information measure used to evaluate semantic information under counter-factual interventions providing a dynamic perspective on the information flow between the system and environment \cite{kolchinsky_semantic_2018, brand_semantic_2024, maurizio2, Sowinski2023}. An alternative common approach is the mutual information between the system and environment \cite{kolchinsky_semantic_2018, ruzzante_synthetic_2023, sowinski2024}. 
Since semantic information is that part of the syntactic information relevant for the system to maintain its viability, and information theory provides various measures of the syntactic information shared between the system and its environment, there is flexibility in defining semantic information relative to other syntactic information measures under counter-factual interventions.
The choice of a syntactic information measure should be guided by its ability to accurately capture the dynamics of the system under consideration.

As the system under study is a (molecular) communication system, where the time-varying concentration profile of particles at the receiver serves as the fundamental signal measurement, our goal is to provide semantic information results that capture the upper limits on communication capacity using particle ensembles.
Building on the semantic information measures employed in \cite{kolchinsky_semantic_2018, ruzzante_synthetic_2023, brand_semantic_2024}, we propose maximal mutual information, i.e., channel capacity as a sensible measure of the interaction between the system and environment to meet the requirement of our analysis.
This allows us to integrate goal-directed constraints into the definition of semantic information while establishing the upper bound on syntactic information transfer.
This approach also entails a system model analysis under the optimal input distribution that maximises $I(X;Y)$, ensuring that the semantic information is not biased by suboptimal inputs.

To quantify the correlation between the system and its environment, we derive the channel capacity measure of the general PIC channel as a function of the symbol interval, the particle release rate at the Tx, and the detection probability at the Rx. Our considered system model accounts for imperfections at the transmitter, the channel, and the receiver through stochastic particle release, diffusion, degradation, and reversible binding to the receiver receptors \cite{pic}.
We assume that the duration of the symbol interval is long enough such that particles from one symbol have a negligible effect on future symbols. We also consider that particles diffuse independently of each other and are detected independently of each other. These assumptions ensure that the channel is memoryless and simplify the channel capacity analysis. 

If $\mathcal{F}$ is the set of all valid input probability density functions, then the capacity of the channel in \eqref{eq: binomial_distributon} as a function of the detection probability $P_{\mathrm{i}}(\tau \vert r_\mathrm{0})$, having a unit of $[\mathcal{C}] =$ bit per second, is defined as:

\begin{equation}\label{e:C}
    \mathcal{C}(\tau \vert r_\mathrm{0}) = \underset{f_\mathrm{X}(p) \in \mathcal{F}}{\mathrm{max}} \dfrac{I(X;Y \vert P_{\mathrm{i}}(\tau \vert r_\mathrm{0}))}{\tau}.
\end{equation}
For the binary input, the PIC is equivalent to a Z channel \cite[Theorem 4]{Farsad2020},\cite{Tallini2002}. 
We define the particle release probability as $p_\mathrm{1}$, $p_\mathrm{0} = 1 - p_\mathrm{1}$, and the 1 to 0 crossover/error probability as:
\begin{equation}\label{e:V}
    \mu_\mathrm{p}(\lambda, \tau \vert r_\mathrm{0}) = (1 - P_{\mathrm{i}}(\tau \vert r_\mathrm{0}))^{N(\lambda, \tau)}.
\end{equation}
\noindent
In the following, we use a short notation for $\mu_\mathrm{p}$ by omitting its explicit dependence on the particle release rate, symbol interval, and space. 
Then, the mutual information from \eqref{e:C} is \cite{MacKay2003}:
\begin{eqnarray}\label{e:I}
    I(X;Y) &=& H(Y) - H(Y\vert X) \nonumber \\
           &=& H_2(p_\mathrm{1}(1 - \mu_\mathrm{p})) - (p_\mathrm{0}H_\mathrm{2}(0) + p_\mathrm{1}H_\mathrm{2}(\mu_\mathrm{p})) \nonumber \\
           &=& H_\mathrm{2}(p_\mathrm{1}(1 - \mu_\mathrm{p})) - p_\mathrm{1}H_\mathrm{2}(\mu_\mathrm{p}),
\end{eqnarray}
where $H(\cdot)$ and $H_\mathrm{2}(\cdot)$ are the general Shannon entropy and binary entropy functions. When $I(X;Y)$ is differentiated with respect to $p_\mathrm{1}$, we obtain:

\begin{equation}\label{e:d}
    \dfrac{\mathrm{d}I(X;Y)}{\mathrm{d}p_\mathrm{1}} = (1 - \mu_\mathrm{p})\mathrm{log_\mathrm{2}} \Biggr[ \dfrac{1 - p_\mathrm{1}(1 - \mu_\mathrm{p})}{p_\mathrm{1}(1 - \mu_\mathrm{p})} \Biggr] - H_\mathrm{2}(\mu_\mathrm{p}).
\end{equation}
Setting this derivative to zero, we obtain the optimal input distribution:

\begin{equation}\label{e:x1*}
    p_\mathrm{1}^{*} = \dfrac{1/(1 - \mu_\mathrm{p})}{1 + 2^{H_\mathrm{2}(\mu_\mathrm{p})/(1 - \mu_\mathrm{p})}}.
\end{equation}
Substituting \eqref{e:x1*} into \eqref{e:I} and using \eqref{e:C}, the channel capacity formula for the binary input PIC is:

\begin{equation}\label{e:cc}
    \mathcal{C}(\tau \vert r_\mathrm{0}) = \dfrac{1}{\tau} \mathrm{log}_\mathrm{2} \Biggr[1 + (1 - \mu_\mathrm{p})\mu_\mathrm{p}^{\dfrac{\mu_\mathrm{p}}{1-\mu_\mathrm{p}}} \Biggr].
\end{equation}
In the following, we use a notation for $\mathcal{C}(\tau)$ by omitting its explicit dependence on space.
Since DDS performance depends on the environment conditions, there is a direct quantitative relationship between channel capacity and functional outcomes (viability).
Therefore, we derive the semantic information at time $\tau$, $S_\epsilon(\tau)$, as the minimum channel capacity among the calculated set of counter-factual interventions $\mathcal{I}$ for which $V_\mathrm{i}(\tau)$, $\mathrm{i} \in \mathcal{I}$ reaches the minimum achievable viability, expressed as:
\begin{equation}\label{e:si}
    S_\epsilon(\tau) = \underset{\mathrm{i} \in \mathcal{I}}{\mathrm{min}} \left\{\mathcal{C}_\mathrm{i}(\tau)\vert V_\mathrm{i}(\tau) \leq \underset{\mathrm{i} \in \mathcal{I}}{\min} V(\tau) + \epsilon\right\}.
\end{equation}
In this formulation, we include a slight increase in viability, $\epsilon \geq 0$, relative to $V(\tau)$, representing variations in viability that are practically negligible (adopted from \cite{brand_semantic_2024}). The derived semantic information refers to that part of the overall channel capacity that influences the DDS viability \cite{bartlett2025}.

\section{Numerical Results \& Discussion}

\begin{table}
    \centering
    \caption{Default parameters for the scenario under investigation.}
    \resizebox{\columnwidth}{!}{%
    \begin{tabular}{lcl}
        \toprule
        \textbf{Parameter} & \textbf{Symbol} & \textbf{Value} \\ 
        \toprule
        Timescale & $\tau$ & $\SI{20}{m \second}$ \\
        Particle release rate & $\lambda$ & $\SI{1000}{particle\,\second^{-1}}$ \\
        Degradation rate & $k_\mathrm{d}$ & $\SI{2e4}{\per \second}$ \\
        Binding rate & $k_\mathrm{f}$ & $\SI{1e-14}{particle^{-1} \meter^3 \second^{-1}}$\\
        Unbinding rate & $k_\mathrm{b}$ & $\SI{2e4}{\per \second}$ \\
        Internalisation rate & $k_\mathrm{i}$ & $\SI{1e3}{\per \second}$\\
        Diffusion coefficient & $D$ & $\SI{5e-9}{\meter^{2} \second^{-1}}$ \\
        Cancer cell radius & $a$ & $\SI{0.5}{\micro \meter}$\\
        Distance between Tx and Rx & $r_\mathrm{0}$ & $\SI{1}{\micro \meter}$ \\
        50\% viability reduction & $C_{\mathrm{th}}$ & $\SI{0.05}{particle\, cell^{-1}}$\\
        Hill coefficient & $n$ & 10 \\
        Tolerated variations in viability & $\epsilon$ & $10^{-2}$ \\
        \bottomrule
    \end{tabular}
    }
    \label{tab:tab}
\end{table}

In this section, we present numerical results for the proposed semantic information framework applied to the considered DDS model (see Sections \ref{sec:2}-\ref{sec:3}). 
We explore whether the DDS optimisation, under the given system configuration (defined by the default parameter set) and constraints, can make the drug delivery process more cost-effective. Using counter-factual interventions, we assess the extent to which drug dosage, DDS targeting or delivery specificity could be adjusted to maximise therapeutic outcomes, i.e., to minimise the viability of the targeted cancer cell.
Based on the DDS model formulation in Sec. \ref{sec:2}, we consider counter-factual interventions by individually altering the following system parameters: particle release rate $\lambda$, degradation rate $k_\mathrm{d}$, binding rate $k_\mathrm{f}$, unbinding rate $k_\mathrm{b}$, and internalisation rate $k_\mathrm{i}$, since they can be modified to some extent during drug characterisation and production.
Unless stated otherwise, Table \ref{tab:tab} lists the default parameter values adopted from \cite{pic} or selected to capture a broad range of potential system responses.

\begin{figure}
    \centering
    \includegraphics[scale=0.5]{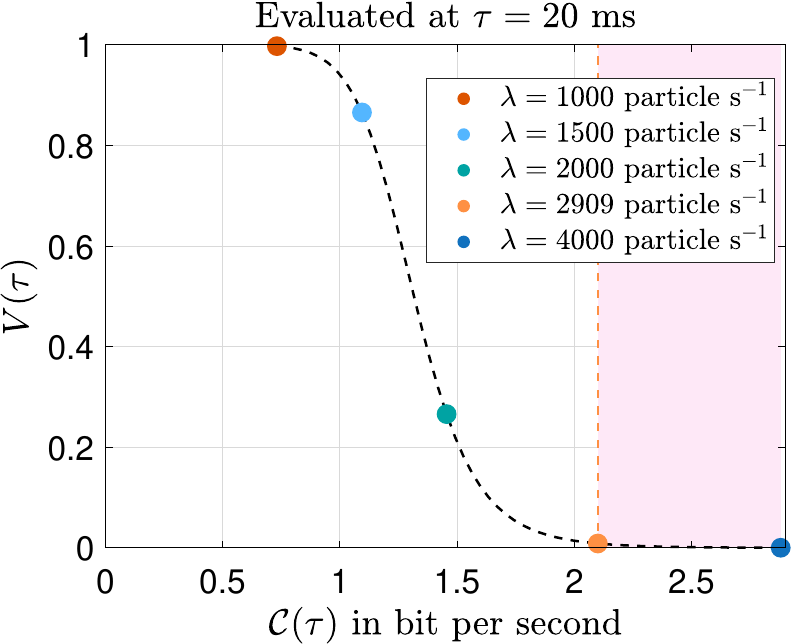}
    \caption{Channel capacity versus viability at $\tau = \SI{20}{m \second}$ for the particle release rate intervention range $\lambda \in \left[1000, 4000\right]$. The vertical dashed orange line marks the derived semantic information $S_{\epsilon}(\tau) = \SI{2.135}{bit\,\second^{-1}}$. The pink-shaded area highlights channel capacities within the observed $\lambda$ intervention range that yield (nearly) constant viability, constrained by $\epsilon$.}
    \label{fig:ev}
\end{figure}

\begin{figure}
    \centering
    \includegraphics[scale=0.5]{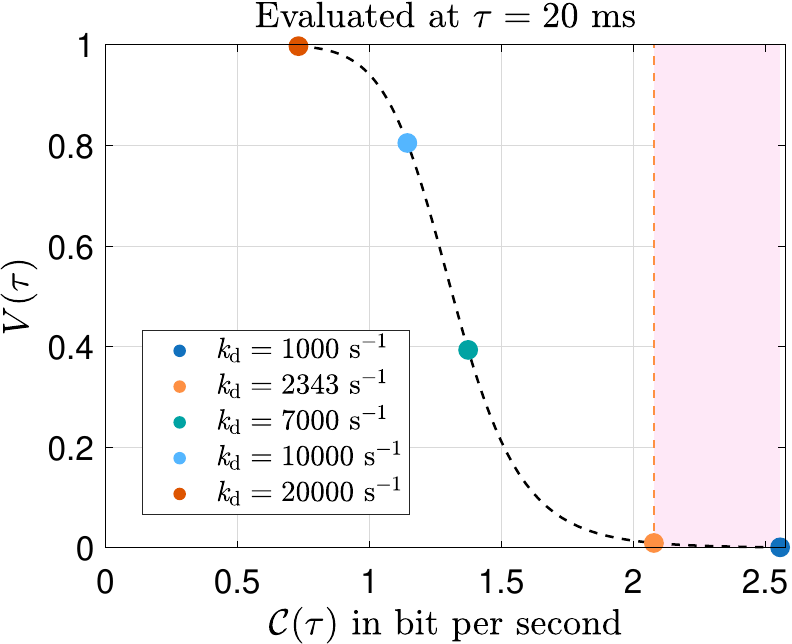}
    \caption{Channel capacity versus viability at $\tau = \SI{20}{m \second}$ for the degradation rate intervention range $k_\mathrm{d} \in \left[1000, 20000\right] \si{\per \second}$. The vertical dashed orange line marks the derived semantic information $S_{\epsilon}(\tau) = \SI{2.126}{bit\,\second^{-1}}$. The pink-shaded area highlights channel capacities within the observed $k_\mathrm{d}$ intervention range that yield (nearly) constant viability, constrained by $\epsilon$.}
    \label{fig:kd2}
\end{figure}

\begin{figure}
    \centering
    \includegraphics[scale=0.5]{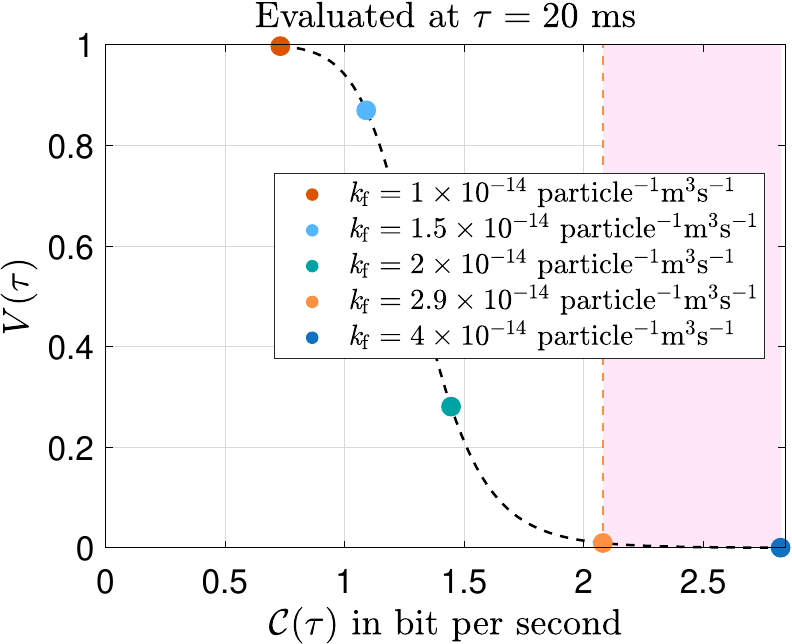} 
    \caption{Channel capacity versus viability at $\tau = \SI{20}{m \second}$ for the binding rate intervention range $k_\mathrm{f} \in \left[1, 4\right] \times 10^{-14} \,\si{particle^{-1}\meter^3\second^{-1}}$. The vertical dashed orange line marks the derived semantic information $S_{\epsilon}(\tau) = \SI{2.071}{bit\,\second^{-1}}$. The pink-shaded area highlights channel capacities within the observed $k_\mathrm{f}$ intervention range that yield (nearly) constant viability, constrained by $\epsilon$.}
    \label{fig:kf2}
\end{figure}

\begin{figure}
    \centering
    \includegraphics[scale=0.5]{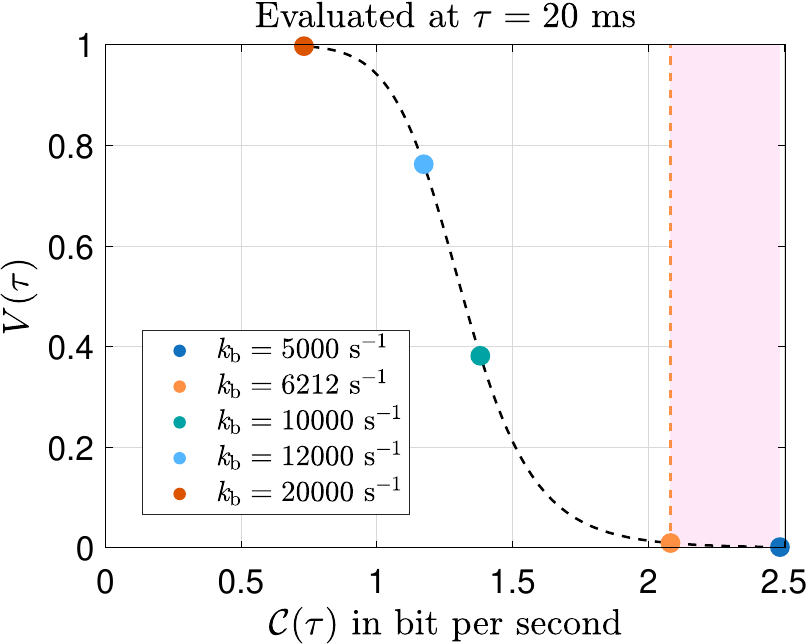}
    \caption{Channel capacity versus viability at $\tau = \SI{20}{m \second}$ for the unbinding rate intervention range $k_\mathrm{b} \in \left[5000, 20000\right] \si{\per \second}$. The vertical dashed orange line marks the derived semantic information $S_{\epsilon}(\tau) = \SI{2.077}{bit\,\second^{-1}}$. The pink-shaded area highlights channel capacities within the observed $k_\mathrm{b}$ intervention range that yield (nearly) constant viability, constrained by $\epsilon$.}
    \label{fig:kb2}
\end{figure}

\begin{figure}
    \centering
    \includegraphics[scale=0.5]{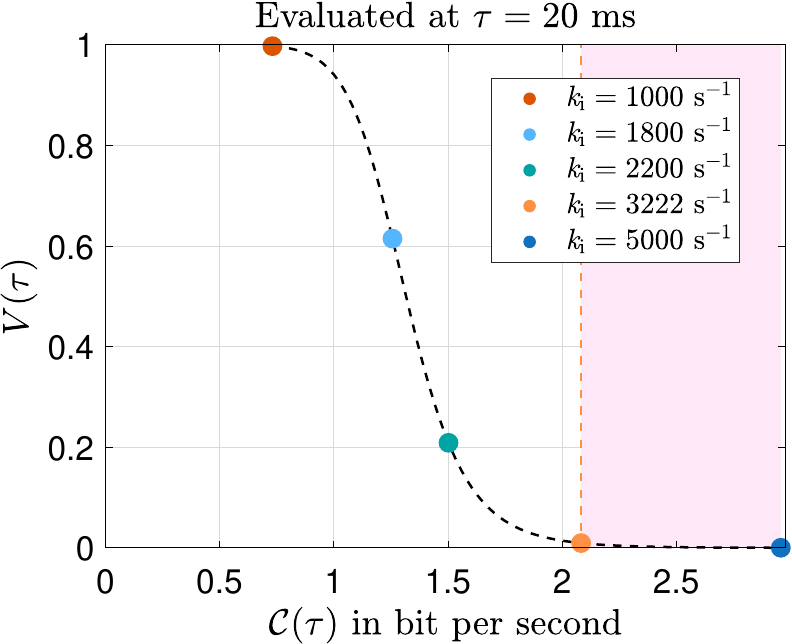}
    \caption{Channel capacity versus viability at $\tau = \SI{20}{m \second}$ for the internalisation rate intervention range $k_\mathrm{i} \in \left[1000, 5000\right] \si{\per \second}$. The vertical dashed orange line marks the derived semantic information $S_{\epsilon}(\tau) = \SI{2.072}{bit\,\second^{-1}}$. The pink-shaded area highlights channel capacities within the observed $k_\mathrm{i}$ intervention range that yield (nearly) constant viability, constrained by $\epsilon$.}
    \label{fig:ki2}
\end{figure}

For each intervention, we first numerically estimate $P_\mathrm{i}(\tau \vert r_0)$ \cite{pic} at $\tau = \SI{20}{m \second}$ using the default values for the remaining DDS parameters. We then compute the viability function $V(\tau)$ in \eqref{e:V1}, and the channel capacity $\mathcal{C}(\tau)$ in \eqref{e:cc}, aiming to generate a viability/channel capacity curve for each intervention type. Finally, we apply \eqref{e:si} to derive semantic information $S_\epsilon$ for the considered interventions.
To address whether DDS optimisation under the default system configuration and constraints enhances cost-effectiveness, we plot in Figs. \ref{fig:ev}-\ref{fig:ki2} interaction between the viability function $V(\tau)$ and the channel capacity $\mathcal{C}(\tau)$, computed at $\tau = \SI{20}{ m \second}$, for various counter-factual interventions.

For the intervention on particle release rate (Fig. \ref{fig:ev}), we consider the range $\lambda \in \left[1000, 4000\right] \si{particle\,\second^{-1}}$. As expected, higher particle release rates reduce cancer cell viability. 
However, a 37\% increase in $\lambda$ (from $2909$ to $\SI{4000}{particle\,\second^{-1}}$) results in only a marginal viability decrease ($\Delta V = \epsilon$).
It is essential to consider that the extent to which $V(\tau)$ changes in response to modifications in the particle release rate directly depends on the characteristics of the PIC system. For a given set of DDS parameters, the viability function $V(\tau)$ of the target cancer cell may reach a plateau region, meaning the cell is unable to internalise additional particles, and excess particles no longer contribute to the therapeutic outcome. This suggests that not all information (or concentration) gathered by the target cancer cell is necessarily relevant to its death, motivating further investigation into semantic information.

The viability plateaus near zero even as $\lambda$ increases. Conversely, once $\lambda$ is sufficiently reduced, the viability rises sharply. 
The point where viability starts to have a rapid increase (observing from right to left) or a plateau (observing from left to right) represents the minimum required channel capacity for the DDS to achieve its therapeutic goal. This channel capacity threshold of dynamical exchange between the system and environment, marked by the orange dashed line in Fig. \ref{fig:ev}, defines the semantic information $S_{\epsilon}(\tau)$, setting a lower bound on $\lambda$ that is required for achieving effective treatment. Based on Fig. \ref{fig:ev}, this critical value is approximately $\lambda \approx \SI{2909}{particle\,\second^{-1}}$, with a corresponding semantic information of $S_{\epsilon}(\tau) = \SI{2.135}{bit\,\second^{-1}}$.

Variations in channel capacity across different $\lambda$ interventions indicate that these interventions affect not only DDS performance but also interaction between the system and its environment. The observed negative correlation between channel capacity and viability confirms that channel capacity, as a semantic information measure, effectively quantifies the DDS's ability to achieve its therapeutic goal (effectiveness level) in a meaningful way.

Numerical results in Figs. \ref{fig:kd2}-\ref{fig:ki2} show that other parameter interventions, within the ranges specified in Table \ref{tab:par},
also lead to rapid plateaus in viability as the parameters are sufficiently modified. This defines the critical parameter values, summarised in Table \ref{tab:par}, that are necessary for the DDS to achieve its therapeutic goal. 
This point also represents the derived semantic information $S_{\epsilon}(\tau)$, i.e., the channel capacity threshold, highlighted by the vertical dashed orange line in Figs. \ref{fig:kd2}-\ref{fig:ki2}, as follows: $\{S_{\epsilon}^{k_\mathrm{d}}(\tau) = 2.126$, $S_{\epsilon}^{k_\mathrm{f}}(\tau) = 2.071$, $S_{\epsilon}^{k_\mathrm{b}}(\tau) = 2.077$, and $S_{\epsilon}^{k_\mathrm{i}}(\tau) = 2.072\} \, \si{bit \, \second^{-1}}$.

The viability curve spans its full possible range, $V(\tau) \in \left[0, 1\right]$, in the demonstrated examples (Figs. \ref{fig:ev}-\ref{fig:ki2}). However, it depends on the timescale $\tau$ and the observed parameter range.


\begin{table}
    \centering
    \caption{Intervention ranges and critical values for the considered DDS parameters.}
    \begin{tabular}{cll}
        \toprule
        Parameter & Intervention Range & Critical Value \\
        \toprule
        $\lambda$ & $\left[1000, 4000\right]$ & $\SI{2909}{particle\,\second^{-1}}$ \\
        $k_\mathrm{d}$ & $\left[1000, 20000\right]$ & $\SI{2343}{\per \second}$\\
        $k_\mathrm{f}$ & $\left[1, 4\right] \times \SI{e-14}{}$ & $\SI{2.9e-14}{particle^{-1} \meter^3 \second^{-1}}$\\
        $k_\mathrm{b}$ & $\left[5000, 20000\right]$ & $\SI{6212}{\per \second}$\\
        $k_\mathrm{i}$ & $\left[1000, 5000\right]$ & $\SI{3222}{\per \second}$\\
        \bottomrule
    \end{tabular}
    \label{tab:par}
\end{table}

After deriving the semantic information, we are able to detect meaningful and meaningless portions within the value ranges of DDS parameters. 
The pink-shaded area in Figs. \ref{fig:ev}-\ref{fig:ki2} highlights the meaningless information, i.e., the channel capacity within the observed intervention range that results in (nearly) constant viability (limited by $\epsilon$). 
This reveals how much a DDS parameter can be intervened without increasing the caner cell's chance of survival, i.e., without decreasing DDS performance. 

\begin{figure}
    \centering
    \includegraphics[scale=0.5]{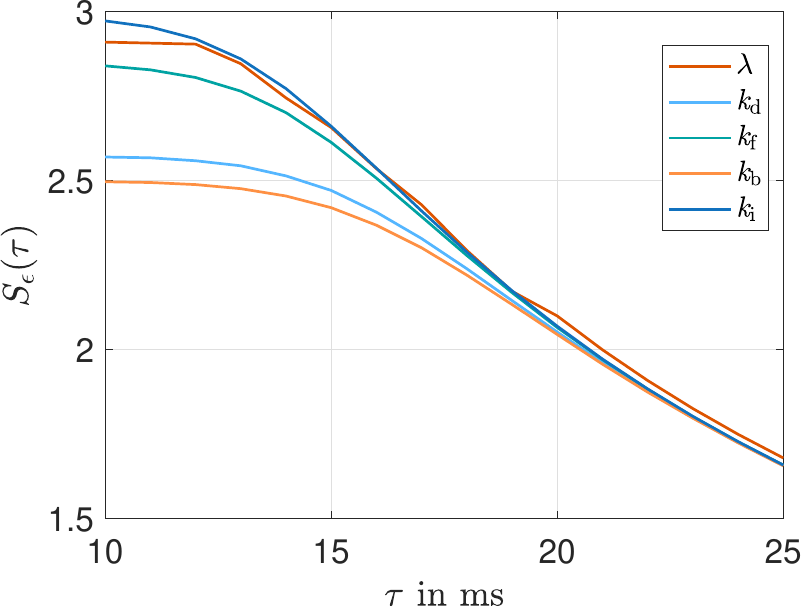}
    \caption{Temporal dynamics of semantic information $S_{\epsilon}(\tau)$ over the time interval $10$ to $\SI{25}{m \second}$ for the counter-factual interventions: particle release rate $\lambda$ (red line), degradation rate $k_\mathrm{d}$ (light blue line), binding rate $k_\mathrm{f}$ (green line), unbinding rate $k_\mathrm{b}$ (orange line), and internalisation rate $k_\mathrm{i}$ (blue line), with intervention ranges specified in Table \ref{tab:par}.}
    \label{fig:si}
\end{figure}

We confirm that the derived semantic information is strictly dependent on the time instant $\tau$ at which the viability function and channel capacity are computed, as outlined in Sec. \ref{sec:3a}. 
Fig. \ref{fig:si} illustrates the temporal evolution of semantic information $S_{\epsilon}(\tau)$, calculated using the considered MC-based DDS model (see Sec. \ref{sec:2}) and the proposed semantic information framework (see Sec. \ref{sec:3}) over the time interval from $10$ to $\SI{25}{m \second}$.
The figure presents results for various counter-factual interventions applied to a DDS: particle release rate $\lambda$ (red line), degradation rate $k_\mathrm{d}$ (light blue line), binding rate $k_\mathrm{f}$ (green line), unbinding rate $k_\mathrm{b}$ (orange line), and internalisation rate $k_\mathrm{i}$ (blue line), with intervention ranges specified in Table \ref{tab:par}. 
The differences in curves suggest that each intervention type uniquely influences the system's information transfer. However, all interventions eventually follow a similar decreasing trend with minor differences. Given the similar trend across the curves, we conclude that the considered DDS parameters contribute to DDS performance, suggesting that DDS optimisation is feasible under the considered constraints.

Therefore, this preliminary study demonstrates the potential for optimising a subset of DDS parameters and designing a cost-effective system by adjusting the selected DDS design parameters $(\lambda, k_\mathrm{d}, k_\mathrm{f}, k_\mathrm{b}, k_\mathrm{i})$ under different criteria and evaluating the semantic information relevant to therapeutic outcomes based on the viability function and channel capacity.
However, the proposed framework may face limitations due to the inherent variability of biological systems (e.g., patient-to-patient differences, caner heterogeneity, immune responses). For instance, optimising and designing a DDS using semantic information theory may require solving complex, high-dimensional models, which can be computationally expensive.
Despite these challenges, the semantic information framework enhances computational dose-response models \cite{DARYAEE2019, gardner2000, tzafriri2009, an2005, gennemark2007, gennemark2024} by shifting the focus from raw drug concentrations to goal-directed information transfer, ensuring that the delivered drug effectively contributes to therapeutic outcomes. Specifically, rather than merely modelling how much drug is present at a target site (a concentration-based approach), the semantic information framework quantifies how much of the drug contributes to a meaningful therapeutic effect (e.g., cancer cell death, immune activation).

\section{Conclusion}

In this paper, we explore the potential of semantic information theory in drug delivery design and optimisation. 
Building on the framework introduced in \cite{kolchinsky_semantic_2018}, we develop a semantic information framework to obtain an information theoretic characterisation of the MC-based computational drug delivery model \cite{pic}. 
Our results demonstrate how syntactic and semantic information measures can be leveraged to guide the DDS design and optimisation under constraints. Specifically, the proposed semantic information measure, based on the channel capacity, quantifies how much system parameters can be intervened without significantly compromising DDS performance, or how much system parameter intervention is required for a DDS to achieve its therapeutic goal.

We argue that applying semantic information theory to DDS models offers a novel perspective capturing coupled system dynamics --- such as those in living systems --- that may be overlooked by purely physical or chemical models. 
Given the framework's ability to quantify the relevance and impact of information on the DDS's ability to achieve therapeutic goals, it has the potential to change the drug delivery design and optimisation by focusing on the meaningful aspects of drug-target interactions. This would implicitly lead to minimising collateral effects while maximising therapeutic performance.

As motivated in this work, this approach can enable DDS design and optimisation across different drugs since each drug type has unique targeting and delivery specificity reflected in the modelled DDS parameters.
Based on the presented results, the semantic information theory can also guide drug design, cell and particle engineering, and resource allocation for externally triggered drug delivery and treatment.
As a future work, we argue that the semantic information approach allows for personalised drug delivery by dynamically adjusting treatment based on individualised patient responses, rather than relying on population-based dosing models.

\bibliographystyle{IEEEtran}
\bibliography{refs}

\end{document}